\documentclass[pra,preprint,superscriptaddress,floatfix,showpacs]{revtex4-1}

\usepackage{amsmath,amsfonts,amssymb}
\usepackage{graphicx}

\def\beq{\begin{equation}}
\def\eeq{\end{equation}}
\def\beqn{\begin{eqnarray}}
\def\eeqn{\end{eqnarray}}

\def\bi {\mbox{\boldmath $i$}}
\def\bo {\mbox{\boldmath $1$}}
\def\bl {\mbox{\boldmath $[$}}
\def\br {\mbox{\boldmath $]$}}
\def\r {{\bf r}}
\def\x {{\bf x}}
\def\y {{\bf y}}

\def\Q {{\bf Q}}
\def\P {{\bf P}}
\def\R {{\bf R}}

\def\X {{\bf X}}
\def\Y {{\bf Y}}

\begin{document}

\title{Solvable Model of a Mixture of Bose-Einstein Condensates}
\author{Shachar Klaiman}
\email{shachar.klaiman@pci.uni-heidelberg.de}
\affiliation{Theoretische Chemie, Physikalisch--Chemisches Institut, Universit\"at Heidelberg, 
Im Neuenheimer Feld 229, D-69120 Heidelberg, Germany}
\author{Alexej I. Streltsov}
\email{alexej.streltsov@pci.uni-heidelberg.de}
\affiliation{Theoretische Chemie, Physikalisch--Chemisches Institut, Universit\"at Heidelberg, 
Im Neuenheimer Feld 229, D-69120 Heidelberg, Germany}
\author{Ofir E. Alon}
\email{ofir@research.haifa.ac.il}
\affiliation{Department of Physics, University of Haifa at Oranim, Tivon 36006, Israel}
\date{\today}

\begin{abstract}
A mixture of two kinds of identical bosons held in a harmonic potential and interacting 
by harmonic particle-particle interactions is discussed. 
This is an exactly-solvable model of a mixture of two trapped
Bose-Einstein condensates 
which allows us to examine analytically various properties. 
Generalizing the treatment in [Cohen and Lee, J. Math. Phys. {\bf 26}, 3105 (1985)], 
closed form expressions for the ground-state energy, wave-function, 
and lowest-order densities are obtained and analyzed for attractive and repulsive intra-species and inter-species particle-particle interactions. 
A particular mean-field solution of the corresponding Gross-Pitaevskii theory is also found analytically. 
This allows us to compare properties of the mixture at the exact, many-body and mean-field levels, 
both for finite systems and at the limit of an infinite number of particles. 
We hereby prove that the exact ground-state energy and lowest-order intra-species and inter-species densities converge at the infinite-particle limit 
(when the products of the number of particles times the intra-species and inter-species
interaction strengths are held fixed)
to the results of the Gross-Pitaevskii theory for the mixture. 
Finally and on the other end, the separability of the mixture's center-of-mass coordinate 
is used to show that the Gross-Pitaevskii theory for mixtures is unable to describe the 
variance of many-particle operators in the mixture, even in the infinite-particle limit. 
Our analytical results show that many-body correlations exist in a mixture of
Bose-Einstein condensates made of any number of particles. 
Implications are briefly discussed.
\end{abstract}


\maketitle 

\section{Introduction}\label{INTRO}

Mixtures of Bose-Einstein condensates,
following their first experimental realizations in ultra-cold atoms \cite{Be1,Be2,Be3,Be4},
have been immensely explored theoretically
for their ground state and excitations, statics and out-of-equilibrium dynamics,
at zero
and finite temperatures, when miscible or immiscible,
and in traps of various shapes and topologies, see, e.g., \cite{Bt1,Bt2,Bt3,Bt4,Bt5,Bt6,Bt7,Bt8,Bt9,Bt10,Bt11,Bt12,Bt13,Bt14,Bt15,Bt16,Bt17,Bt18,Bt19,Bt20,
Bt21,Bt22,Bt23,Bt24,Bt25,Bt26,Bt27,Bt28,Bt30,Bt30h,Bt31,Bt32,Bt33,Bt34,Bt36,Bt37,Bt38,Bt39,Bt40,
Bt41,Bt42,Bt42h,Bt43,Bt44,Bt45,Bt46,Bt47,Bt48},
representing in our viewing how fascinating and rich these quantum systems are.
Methodologically, bosonic mixtures have been treated
by a variety of many-body approaches and numerical tools, 
as well as within Gross-Pitaevskii, mean-field theory.
Whereas the Gross-Pitaevskii theory is often employed, 
only the use of many-body theory can actually ensure when the mean-field 
theory provides an adequate description and when it is not. 
Obviously and quite generally, a many-body description is more demanding than a mean-field description. 
In this respect, having an exactly-solvable many-body model at hand is, of course, the ideal situation.

In this work 
we present such an exactly-solvable model of a mixture of two trapped Bose-Einstein condensates. 
Explicitly, we consider two kinds of identical bosons held in a harmonic potential and interacting 
by harmonic particle-particle interactions, or, briefly, the harmonic-interaction model (HIM)
for bosonic mixtures. 
We emphasize that the HIM model for single-species bosons is well known 
and has 
provided ample many-body, exact results as well as means
to benchmark numerical investigations 
\cite{HIM_Cohen,HIM_Po0,HIM_Yan,HIM_Po1,HIM_Benchmarks,LR_MCTDHB_Bench}.
Similarly, the HIM model for fermions has been employed, see, e.g., \cite{HIM_Po0,Schilling_FER_HIM_WORK,Axel_MCTDHF_HIM}.
We would also like to mention the use of an analytical treatment based on 
coupled harmonic oscillators for vibrations \cite{Lenz_Domcke_Harmonic}, 
which are distinguishable degrees of freedom. 
 
The mixture's HIM model to be derived below 
allows us to examine analytically various properties, 
in particular the ground-state energy, wave-function, 
and lowest-order intra-species and inter-species densities, 
and how they depend on the number of particles and interactions in the mixture. 
A plethora of exact, many-body results is reported and analyzed.

In addition, 
we also find analytically the ground-state 
solution of the Gross-Pitaevskii theory for the mixture's HIM.
This enables us to compare properties of the mixture at the exact, many-body and mean-field levels, 
both for finite mixtures and
in the limit of an infinite number of particles. 
The later topic,
i.e., when the many-body and mean-field descriptions of a Bose-Einstein condensate coincide,
is of much interest for single-species bosons 
\cite{Yngvason_PRA,Lieb_PRL,Erdos_PRL,MATH_ERDOS_REF,Variance,TD_Variance}, 
but has yet to be addressed for mixtures.
We hereby prove that the exact ground-state energy and lowest-order intra-species 
and inter-species densities converge at the infinite-particle limit 
to the results of the Gross-Pitaevskii theory for the mixture.

Last but not least, 
the separability of the mixture's center-of-mass coordinate 
is used to show that the Gross-Pitaevskii theory for mixtures
can deviate strongly when computing 
the variance of many-particle operators in the mixture, 
even in the infinite-particle limit.
Unlike the variance of operators of a single particle,
which is a basic notion 
in any quantum mechanics textbook \cite{QM_book},
the variance of operators of many-particle systems is more involved \cite{Variance,TD_Variance},
also see \cite{Oriol_Robin} in this context.
Our analytical results show that many-body correlations exist in a trapped mixture of 
Bose-Einstein condensates consisting of 
{\it any number of particles}.

\section{The Harmonic-Interaction Model for Mixtures}\label{HIM_MIX}

Consider a mixture of two kinds of identical bosons which we denote $A$ and $B$.
The bosons are trapped in a three-dimensional isotropic harmonic potential and interact between them
via harmonic particle-particle interactions.
We focus in this work on what might be considered the simplest case, the symmetric mixture.
This is a mixture consisting 
of $M$ bosons of type $A$ and an equal number of $M$ bosons of type $B$,
all having the same mass (taken below to be $1$) and trapped in the same harmonic potential (of frequency $\omega$).
The total number of particles is denoted by $N=2M$.
Furthermore, the two intra-species interactions are alike (denoted by $\lambda_1$).
The inter-species interaction is donated by $\lambda_2$.
Positive values of $\lambda_1$ and $\lambda_2$ mean 
attractive particle-particle interactions
whereas negative values imply repulsive interactions.
Of course, the intra-species interactions may be repulsive and the inter-species attractive,
and vice versa. 

The mixture's Hamiltonian is then given by ($\hbar=1$)
\beqn\label{HAM_MIX}
& & \hat H(\x_1,\ldots,\x_M,\y_1,\ldots,\y_M) = 
 \sum_{j=1}^M 
\left[\left( -\frac{1}{2} \frac{\partial^2}{\partial \x_j^2} + \frac{1}{2} \omega^2 \x_j^2 \right) +
\left( -\frac{1}{2} \frac{\partial^2}{\partial \y_j^2} + \frac{1}{2} \omega^2 \y_j^2 \right)\right] + \nonumber \\
& & \qquad + \lambda_1 \sum_{1 \le j < k}^M \left[ (\x_j-\x_k)^2 + (\y_j-\y_k)^2 \right]  
+ \lambda_2 \sum_{j=1}^M \sum_{k=1}^M (\x_j-\y_k)^2.
\eeqn
Here, the coordinates $\x_j$ denote bosons of type $A$ and $\y_k$ bosons of type $B$.
We work in Cartesian coordinates where the vector $\x=(x_1,x_2,x_3)$ denotes an $A$ particle's position in three dimensions,
and $\frac{1}{i}\frac{\partial }{\partial \x}=
\frac{1}{i}\left(\frac{\partial}{\partial x_1},\frac{\partial}{\partial x_2},\frac{\partial}{\partial x_3}\right)$ its momentum.
To avoid cumbersome notation,
we denote $\x^2 \equiv x_1^2+x_2^2+x_3^2$ and
$\frac{\partial^2}{\partial \x^2} \equiv \frac{\partial^2}{\partial x_1^2} + \frac{\partial^2}{\partial x_2^2}    
+ \frac{\partial^2}{\partial x_3^2}$.
Analogous notation is employed for the vector $\y$ of the $B$ species bosons.
In what follows and when needed, we use the vector $\r$ to denote the position of a particle of either kind.
Opening the braces of the particle-particle interaction terms and 
collecting the diagonal contributions together we have
\beqn\label{HAM_MIX_open}
& & \hat H =
 \sum_{j=1}^M 
\left\{ -\frac{1}{2} \left(\frac{\partial^2}{\partial \x_j^2} + \frac{\partial^2}{\partial \y_j^2}\right) +
\frac{1}{2}\left[\omega^2 + (N-2)\lambda_1 + N\lambda_2\right] (\x_j^2 + \y_j^2) \right\} - \nonumber \\ 
& & \qquad - 2\lambda_1 \sum_{1 \le j < k}^M (\x_j \cdot \x_k + \y_j \cdot \y_k) - 
 2\lambda_2 \sum_{j=1}^M \sum_{k=1}^M \x_j \cdot \y_k.
\eeqn
As we shall see below,
it is instrumental to combine
the $j$th coordinate in the $A$ and $B$ species 
together. 

To diagonalize the Hamiltonian (\ref{HAM_MIX}) we build on the single-species HIM 
transformed coordinates \cite{HIM_Cohen} 
(here, for each of the $A$ and $B$ species) 
and make the following coordinate transformation
\beqn\label{MIX_COOR}
& & \Q_k = \frac{1}{\sqrt{k(k+1)}} \sum_{j=1}^{k} (\x_{k+1}-\x_j), \qquad 1 \le k \le M-1,  \nonumber \\
& & \Q_{M-1+k} = \frac{1}{\sqrt{k(k+1)}} \sum_{j=1}^{k} (\y_{k+1}-\y_j), \qquad 1 \le k \le M-1,  \nonumber \\
& & \Q_{N-1} = \frac{1}{\sqrt{N}} \sum_{j=1}^M (\x_j - \y_j), \nonumber \\
& & \Q_N = \frac{1}{\sqrt{N}} \sum_{j=1}^M (\x_j + \y_j). \
\eeqn
The meaning of (\ref{MIX_COOR}) is as follows: 
The first group of $M-1$ coordinates are relative coordinates of the $A$ species;
the second group of $M-1$ coordinates are relative coordinates of the $B$ species;
$\Q_{N-1}$ can be seen as a relative coordinate between the center-of-mass of the $A$ species 
and the center-of-mass of the $B$ species; 
and, finally, $\Q_N$ is (proportional to) the center-of-mass coordinate of all particles in the mixture.
Further discussion of this coordinate transformation is given in the Appendix.
With the coordinate transformation (\ref{MIX_COOR}) the kinetic energy remains diagonal,
\beq\label{MIX_KIN}
\sum_{j=1}^M
\left(\frac{\partial^2}{\partial \x_j^2} + \frac{\partial^2}{\partial \y_j^2}\right) =
\sum_{k=1}^N
\frac{\partial^2}{\partial \Q_k^2},
\eeq
and the following quadratic relations hold
\beqn\label{MIX_POT_1}
& & \sum_{k=1}^{N-2} \Q_k^2 = \left(1-\frac{2}{N}\right) \sum_{j=1}^M (\x_j^2+\y_j^2) 
-\frac{4}{N} \sum_{1 \le j < k}^M (\x_j \cdot \x_k + \y_j \cdot \y_k), \nonumber \\
& & \Q^2_{N-1} = \frac{1}{N} \left\{ \sum_{j=1}^M (\x_j^2+\y_j^2) 
+ 2 \left[ \sum_{1 \le j < k}^M (\x_j \cdot \x_k + \y_j \cdot \y_k)
 - \sum_{j=1}^M \sum_{k=1}^M \x_j \cdot \y_k \right] \right\}, \nonumber \\
& & \Q^2_N = \frac{1}{N} \left\{ \sum_{j=1}^M (\x_j^2+\y_j^2) 
+ 2 \left[ \sum_{1 \le j < k}^M (\x_j \cdot \x_k + \y_j \cdot \y_k)
 + \sum_{j=1}^M \sum_{k=1}^M \x_j \cdot \y_k \right] \right\}. \
\eeqn

Using (\ref{MIX_COOR}), (\ref{MIX_KIN}), and (\ref{MIX_POT_1}), 
the Hamiltonian (\ref{HAM_MIX}) transforms to the diagonal form
\beqn\label{HAM_DIAG}
& & \hat H(\Q_1,\ldots,\Q_N) = 
\sum_{k=1}^{N-2} \left( -\frac{1}{2} \frac{\partial^2}{\partial \Q_k^2} + \frac{1}{2} \Omega_{rel}^2 \Q_k^2 \right) + 
\left(-\frac{1}{2} \frac{\partial^2}{\partial \Q_{N-1}^2} + \frac{1}{2} \Omega_{AB}^2 \Q_{N-1}^2\right) + \nonumber \\
& & \qquad \qquad +\left(-\frac{1}{2} \frac{\partial^2}{\partial \Q_N^2} + \frac{1}{2} w^2 \Q_N^2\right). \ 
\eeqn
The mixture's transformed Hamiltonian (\ref{HAM_DIAG}) is that of $N$ uncoupled harmonic oscillators
with the three frequencies
\beq\label{MIX_FREQ}
  \Omega_{rel} = \sqrt{\omega^2 + N(\lambda_1+\lambda_2)}, \qquad
  \Omega_{AB} = \sqrt{\omega^2 + 2N\lambda_2}, \qquad
 \omega.
\eeq
The multiplicity of the frequencies is $N-2$, $1$, and $1$, respectively.

Let us examine the frequencies (\ref{MIX_FREQ}) and their dependence on the
particle-particle interactions more closely.
The frequency of all intra-species relative coordinates $\Q_1,\ldots,\Q_{2(M-1)}$,
$\Omega_{rel}$, depends on both the intra-species $\lambda_1$ and inter-species $\lambda_2$
interactions, yet only through their sum $\lambda_1+\lambda_2$.
The frequency $\Omega_{AB}$ of the intra-species relative coordinate $\Q_{N-1}$ 
depends on $\lambda_2$ only.
The frequency of the center-of-mass degree of freedom $\Q_N$,
which is equal to the trap frequency $\omega$, 
naturally does not depend on either.
When $\lambda_1=0$,
i.e., when the intra-species interactions vanish,
we are still dealing with a mixture described by three frequencies,
solely due to the inter-species interaction which couples all $N$ particles together.
On the other hand, when $\lambda_1+\lambda_2=0$,
we get $\Omega_{rel}=\omega$ and
the mixture is now described by two frequencies only.
We will analyze this case further below.
When the intra- and inter-species interactions are equal, $\lambda_1=\lambda_2$,
the frequency of the relative coordinate between the center-of-masses of the $A$ and $B$ species 
degenerates to that of the other relative coordinates, 
i.e., $\Omega_{AB}=\Omega_{rel}$,
and the standard, single-species 
HIM model of $N$ bosons is recovered.
Finally, in the limiting case when $\lambda_2=0$,
we have $\Omega_{AB}=\omega$ and, as one would expect, 
we find two independent species
each described by the HIM model 
for $M$ bosons.

The frequencies (\ref{MIX_FREQ}) must be positive in order for a bound solution to exist.
This dictates bounds on both the intra-species $\lambda_1$ and inter-species $\lambda_2$ interactions 
which are:
\beqn\label{BOUNDS}
& & \Omega_{AB}^2=\omega^2+2N\lambda_2>0 \quad \Longrightarrow \quad \lambda_2> -\frac{\omega^2}{2N}, \nonumber \\
& & \Omega_{rel}^2 = \omega^2 + N(\lambda_1+\lambda_2)>0
\quad \Longrightarrow \quad \lambda_1 > - \lambda_2 + \frac{\omega^2}{N}. \
\eeqn
The meaning of these bounds are as follows:
The inter-species interaction $\lambda_2$ is bounded from below,
irrespective of the intra-species interaction $\lambda_1$, 
otherwise the mixture cannot be trapped in the harmonic potential.
On the other hand, the intra-species interaction $\lambda_1$ is limited by the chosen 
inter-species interaction $\lambda_2$.

We can now proceed and 
prescribe the normalized ground-state wave-function
\beq\label{WAVE_FUN_1}
\Psi(\Q_1,\ldots,\Q_N) = 
\left(\frac{\Omega_{rel}}{\pi}\right)^{\frac{3(N-2)}{4}} 
\left(\frac{\Omega_{AB}}{\pi}\right)^{\frac{3}{4}}
\left(\frac{\omega}{\pi}\right)^{\frac{3}{4}}
e^{-\frac{1}{2} \left(\Omega_{rel} \sum_{k=1}^{N-2} \Q_k^2 + \Omega_{AB} \Q_{N-1}^2 + \omega \Q_N^2 \right)},
\eeq
along with the ground-state eigen-energy
\beq\label{HIM_MIX_GS_E}
E = \frac{3}{2} \left[ (N-2) \Omega_{rel} + \Omega_{AB} + \omega \right] =
\frac{3}{2} \left[ (N-2) \sqrt{\omega^2 + N(\lambda_1+\lambda_2)} + \sqrt{\omega^2 + 2N\lambda_2} + \omega \right].
\eeq 
To express the wave-function with respect to the original spatial coordinates we use the relations 
(\ref{MIX_POT_1}) and find
\beqn\label{WAVE_FUN_2}
& & \Psi(\x_1,\ldots,\x_M,\y_1,\ldots,\y_M) = \left(\frac{\Omega_{rel}}{\pi}\right)^{\frac{3(N-2)}{4}} 
\left(\frac{\Omega_{AB}}{\pi}\right)^{\frac{3}{4}} \left(\frac{\omega}{\pi}\right)^{\frac{3}{4}} \times  \\
& & \quad \times 
e^{-\frac{\alpha}{2} \sum_{j=1}^M \x_j^2 - \beta \sum_{1 \le j < k}^M \x_j \cdot \x_k} \times
e^{-\frac{\alpha}{2} \sum_{j=1}^M \y_j^2 - \beta \sum_{1 \le j < k}^M \y_j \cdot \y_k} \times
e^{+\gamma \sum_{j=1}^M \sum_{k=1}^M \x_j \cdot \y_k}, \nonumber \
\eeqn 
where
\beq\label{WAVE_FUN_3}
\alpha=\Omega_{rel} \left(1-\frac{2}{N}\right) + (\Omega_{AB}+\omega) \frac{1}{N}
=\Omega_{rel}+\beta, \quad 
\beta=\frac{1}{N}(-2\Omega_{rel}+\Omega_{AB}+\omega), \quad
\gamma=\frac{1}{N}(\Omega_{AB}-\omega).
\eeq
The mixture's ground-state wave-function (\ref{WAVE_FUN_2}) is
seen to be comprised of a product of an $A$ part, 
an equivalent $B$ part, 
and a coupling $AB$ part.

\section{Properties and Analysis}\label{PROP_ANAL}

The HIM model for a mixture of two Bose-Einstein condensates presented above admits
a wealth of properties that can all be studied analytically.
We present in this section a rather detailed account of the ground-state
energy, intra- and inter-species densities, a mean-field solution,
and the variance of center-of-mass operators in the mixture.
Side by side the plethora of closed-form expressions and results,
a guiding line of our exploration are properties of
finite mixtures as well as at the infinite-particle limit (to be defined precisely below),
and how the many-body and mean-field solutions are related to each other.

\subsection{Ground-state energy}

It is instructive and interesting to analyze the ground-state energy (\ref{HIM_MIX_GS_E}) in several cases.
As mentioned above, 
for $\lambda_1=\lambda_2$ we recover the ground-state energy 
of the single-species HIM model \cite{HIM_Cohen}.
Keeping $\lambda_1$ and $\lambda_2$ fixed and increasing the number of particles $N$
the energy to leading order in $N$ scales like $N^{\frac{3}{2}}$,
i.e. adding more particles increases further the mixture's energy.
This situation applies as long as the bounds (\ref{BOUNDS}) are not reached,
that is when the mixture is predominantly attractive, $\lambda_1+\lambda_2>0$ and $\lambda_2>0$.

An interesting situation occurs for $\lambda_1+\lambda_2=0$, i.e.,
when the intra- and inter-species 
interactions 
are exactly opposite in sign.
In this case, as we touched upon above, 
$\Omega_{rel}$ degenerates to $\omega$,
and the energy (\ref{HIM_MIX_GS_E}) reduces to
\beq\label{HIM_MIX_GS_E0}
E |_{\lambda_1+\lambda_2=0} = \frac{3}{2} \left[ (N-1) \omega + \sqrt{\omega^2 + 2N\lambda_2}\right]
\eeq  
and seen to depend only on the inter-species interaction $\lambda_2$.
Now, the ground-state energy (\ref{HIM_MIX_GS_E0}) is linear to leading order in the number of particles $N$.

Using the bounds for $\lambda_1$ and $\lambda_2$ in (\ref{BOUNDS}),
we obtain that the mixture's energy 
is bound from below by $E > \frac{3}{2}\omega$,
which is obtained for $\Omega_{rel} \to 0^+$ and $\Omega_{AB} \to 0^+$.
This means that all relative degrees of freedom are marginally bound,
and essentially only the center-of-mass degree of freedom is bound in the harmonic trap. 

We now move to analyze the mixture's energy in the so-called infinite-particle limit.
To this end, we introduce the intra-species $\Lambda_1=\lambda_1(M-1)$ and
inter-species $\Lambda_2=\lambda_2M$ interaction parameters.
In the infinite-particle limit the interaction parameters $\Lambda_1$ and $\Lambda_2$ are kept fixed,
while the number of particles is increased.
Hence, the interaction strengths $\lambda_1$ and $\lambda_2$ 
diminish accordingly.
The two interaction parameters $\Lambda_1$ and $\Lambda_2$ appear
naturally in the mean-field treatment discussed below,
and would 
facilitate the comparison between 
the exact, many-body and mean-field solutions
of the mixture.

Keeping $\Lambda_1$ and $\Lambda_2$ constant,
the energy per particle 
in the limit of an infinite number of particles reads
\beq\label{HIM_MIX_GS_E_INF}
\lim_{N\to \infty} \frac{E}{N} = \frac{3}{2}\sqrt{\omega^2 + 2(\Lambda_1+\Lambda_2)}.
\eeq
We see that the limit 
of the energy per particle
depends on the sum of intra- and inter-species interaction parameters $\Lambda_1+\Lambda_2$ only.
Finally, in the particular case when the interaction parameters are fixed and opposite in sign, i.e., 
$\Lambda_1+\Lambda_2=0$,
we find in the infinite-particle limit
\beq\label{HIM_MIX_GS_E0_INF}
\lim_{N\to \infty} \frac{E |_{\Lambda_1+\Lambda_2=0}}{N} = \frac{3}{2}\omega. 
\eeq
The meaning of (\ref{HIM_MIX_GS_E0_INF}) is that the energy per particle of the 
interacting system becomes that of the non-interacting system;
A situation that cannot occur in the single-species system.
Here, the intra- and inter-species interaction parameters `cancel' each other in the infinite-particle limit,
at least as far as the energy per particle in concerned.
We will return to this peculiar situation below 
when analyzing the density and mean-field solution of the mixture.

\subsection{Intra- and inter-species densities}

We start from the $N$-particle density of the mixture
\beqn\label{MB_DENS}
& & |\Psi(\x_1,\ldots,\x_M,\y_1,\ldots,\y_M)|^2 = 
\left(\frac{\Omega_{rel}}{\pi}\right)^{\frac{3(N-2)}{2}} 
\left(\frac{\Omega_{AB}}{\pi}\right)^{\frac{3}{2}} \left(\frac{\omega}{\pi}\right)^{\frac{3}{2}} \times  \\
& & \quad \times 
e^{-\alpha \sum_{j=1}^M \x_j^2 - 2\beta \sum_{1 \le j < k}^M \x_j \cdot \x_k} \times
e^{-\alpha \sum_{j=1}^M \y_j^2 - 2\beta \sum_{1 \le j < k}^M \y_j \cdot \y_k} \times
e^{+2 \gamma \sum_{j=1}^M \sum_{k=1}^M \x_j \cdot \y_k}, \nonumber \
\eeqn 
which for convenience is here normalized to unity.
Let the $AB$ two-body density be
\beq\label{DENS_AB}
\rho_{AB}(\x,\y) = M^2 \int d\x_2 \cdots d\x_M d\y_2 \cdots d\y_M 
|\Psi(\x,\x_2,\ldots,\x_M,\y,\y_2,\ldots,\y_M)|^2.
\eeq
It is the lowest-order 
inter-species density of the mixture.
Then, the intra-species one-body densities are given by
\beq\label{DENS_A_B}
\rho_A(\x) = \frac{1}{M} \int d\y \rho_{AB}(\x,\y) \equiv
\frac{1}{M} \int d\x \rho_{AB}(\x,\y) = \rho_B(\y),
\eeq
and normalized to the number of bosons of each kind, $M$.

To obtain the lowest-order intra- and inter-species densities 
(\ref{DENS_A_B}) and (\ref{DENS_AB}), respectively,  
we need to perform {\it multiple} integrations of the $N$-particle density (\ref{MB_DENS}).
Moreover, the later contains a coupling $AB$ part because of the 
inter-species interaction,
also see below (\ref{WAVE_FUN_2}).
Thus, without an appropriate construction, 
the task becomes quickly impractical with increasing $N$.
Fortunately, we are able to generalize the very useful recurrence integration construction  
for the densities of the single-species HIM put forward in \cite{HIM_Cohen} 
to the mixture's HIM model (\ref{HAM_MIX}).

To integrate the $N$-particle density we begin by introducing the auxiliary function
\beqn\label{FUN_N_HALF_1}
& & F_M(\x_1,\ldots,\x_M,\y_1,\ldots,\y_M;
\alpha,\beta,C_M,D_M) = \nonumber \\
& & = e^{-\alpha\sum_{j=1}^M (\x_j^2+\y_j^2) 
-2\beta\sum_{1 \le j < k}^M (\x_j \cdot \x_k+\y_j \cdot \y_k)
-C_M (\X_M+\Y_M)^2
+2D_M \X_M \cdot \Y_M} = \nonumber \\
& & = e^{-\alpha\sum_{j=1}^M (\x_j^2+\y_j^2) 
-2\beta\sum_{1 \le j < k}^M (\x_j \cdot \y_k+\y_j \cdot \y_k)
-C_M (\X_M^2+\Y_M^2)
+2(D_M-C_M)\X_M \cdot \Y_M}, \
\eeqn  
where $\alpha$, $\beta$, and $\gamma$ are given in (\ref{WAVE_FUN_3}),
\beq\label{FUN_N_HALF_2}
\X_M = \sum_{j=1}^M \x_j, \qquad
\Y_M = \sum_{j=1}^M \y_j,
\eeq
are vectors, 
and $C_M$ and $D_M$ are constants to play 
a further role below.
We note 
that if one were to equate the $N$-body density (\ref{MB_DENS}) 
and the auxiliary function $F_M$ then this would imply 
that $C_M=0$ and $D_M=\gamma$. 
Finally, we note that the following relations
\beqn\label{FUN_N_HALF_3}
& & \X^2_M + \Y^2_M =
\X^2_{M-1} + \Y^2_{M-1} 
+ [\x^2_M + \y^2_M +
2(\x_M \cdot \X_{M-1} + \y_M \cdot \X_{M-1})], \nonumber \\
& & \qquad \X_M \cdot \Y_M = 
\X_{M-1} \cdot \Y_{M-1} +
[\x_M \cdot \Y_{M-1} + \y_M \cdot \X_{M-1} + 
 \x_M \cdot \y_M], \
\eeqn
hold
between the vectors introduced in (\ref{FUN_N_HALF_2}) and the 
pair of variables $\x_M$ and $\y_M$.

The key point is to perform 
the integrations in pairs 
reducing thereby the order of the auxiliary function $F_M$.
The basic ingredient is the double-Gaussian integral 
over $\x_j$ and $\y_j$
\beqn\label{BASIC_DOUBLE_INT}
& & \int d\x_j d\y_j e^{-a(\x_j^2+\y_j^2)-
2[\x_j \cdot (b\X_{j-1}+b'\Y_{j-1}) + \y_j \cdot (b\Y_{j-1}+b'\X_{j-1})] 
+2c\x_j \cdot \y_j} = \nonumber \\
& & \qquad = \frac{\pi^3}{(a^2-c^2)^\frac{3}{2}} 
e^{+p(\X_{j-1}^2+\Y_{j-1}^2)+2q\X_{j-1} \cdot \Y_{j-1}}, \
\eeqn
where $a$, $b$, $b'$, $c$, $p$, and $q$ are scalars connected by the relations
\beq\label{BASIC_DOUBLE_INT_2}
p=\frac{a(b^2+b'^2)+2bb'c}{a^2-c^2}, \quad q=\frac{c(b^2+b'^2)+2abb'}{a^2-c^2}, \quad
(a\mp c)(p\pm q)=(b\pm b')^2.
\eeq
Furthermore, to perform multiple double-integration steps like (\ref{BASIC_DOUBLE_INT}),
we seek for a recurrence relation and thus write
\beqn\label{FUN_N_HALF_M1_1}
& & \int d\x_M d\y_M F_M = \nonumber \\
& & = e^{-\alpha\sum_{j=1}^{M-1} (\x_j^2+\y_j^2) 
-2\beta\sum_{1 \le j < k}^{M-1} (\x_j \cdot \x_k +\y_j \cdot \y_k)
-C_M (\X_{M-1}^2+\Y_{M-1}^2)
+2(D_M-C_M)\X_{M-1} \cdot \Y_{M-1}} \times \nonumber \\
& & \int d\x_M d\y_M e^{-(\alpha+C_M) (\x_M^2+\y_M^2)} \times \nonumber \\
& & e^{-2\{\x_M \cdot [(\beta+C_M)\X_{M-1} - 
(D_M-C_M) \cdot \Y_{M-1}] +
2\{\y_M \cdot [(\beta+C_M)\Y_{M-1} - 
(D_M-C_M) \cdot \X_{M-1}]\} 
+2(D_M-C_M)\x_M \cdot \y_M} = \nonumber \\
& & = \frac{\pi^3}{[(\alpha+D_M)(\alpha+\tilde C_M)]^\frac{3}{2}} 
F_{M-1}(\x_1,\ldots,\x_{M-1},\y_1,\ldots,\y_{M-1};
\alpha,\beta,C_{M-1},D_{M-1}). \
\eeqn
The equality (\ref{FUN_N_HALF_M1_1}) relates the 
auxiliary function $F_M$ with $2M$ coordinates and constants $C_M$
and $D_M$ to the function $F_{M-1}$ of the same functional form,
with $2(M-1)$ coordinates and corresponding constants $C_{M-1}$ and $D_{M-1}$.
To simplify the relation between the constants in $F_M$ and $F_{M-1}$,
it is useful to take appropriate linear combinations.
The final result reads
\beqn\label{FUN_N_HALF_M1_2}
& & \tilde C_{M-1} = \tilde C_M - \frac{(\beta+\tilde C_M)^2}{\alpha+\tilde C_M}, \qquad \tilde C_M = (2C_M-D_M) = -\gamma,
\nonumber \\
& & 
D_{M-1} = D_M - \frac{(\beta+D_M)^2}{\alpha+D_M}, \qquad D_M=\gamma, \
\eeqn 
where the 
values
of $\tilde C_M$ and $D_M$
are obtained
when we equate the 
auxiliary function $F_M$ and the $N$-body density (\ref{MB_DENS}).

We can now write the recurrence relation connecting the function $F_j$ 
with $2j$ variables and constants $C_j$ and $D_j$
to the function $F_{j-1}$ 
with $2(j-1)$ variables and constants $C_{j-1}$ and $D_{j-1}$.
The equality looks like (\ref{FUN_N_HALF_M1_1}) and need not be pasted here.
What is important is the recurrence relation between the corresponding constants which is given by 
\beqn\label{FUN_N_HALF_M1_3}
& & \tilde C_{j-1} = \tilde C_j - \frac{(\beta+\tilde C_j)^2}{\alpha+\tilde C_j}, \qquad \tilde C_j = (2C_j-D_j),
\nonumber \\
& & 
D_{j-1} = D_j - \frac{(\beta+D_j)^2}{\alpha+D_j}. \
\eeqn
In turn, the recursion ends with the `lowest-order' auxiliary function
\beq\label{FUN_N_HALF_END}
F_1(\x_1,\y_1;
\alpha,\beta,C_1,D_1) =  e^{-(\alpha+C_1) (\x_1^2+\y_1^2) 
+2(D_1-C_1)\x_1 \cdot \y_1}.
\eeq
As we shall see, 
the constants $C_1$ and $D_1$ are required to evaluate 
the intra- and inter-species densities of the mixture (\ref{DENS_A_B}) and (\ref{DENS_AB}).
  
Interestingly and importantly, 
the recurrence relations (\ref{FUN_N_HALF_M1_3})
for the parameters $\tilde C_j$ and $D_j$ appearing 
in the double-Gaussian integrations of the mixture's auxiliary function 
have exactly the same structure as the recurrence relation 
emerging in the single-Gaussian integrations of the HIM system \cite{HIM_Cohen}.
Thus, we use this result directly and write for 
the final result of the respective solutions
\beqn\label{FUN_N_HALF_M1_4}
& & \tilde C_j = -\alpha +\frac{(\alpha-\beta)(1+j \eta_-)}{1+(j+1)\eta_-}, \qquad
D_j = -\alpha +\frac{(\alpha-\beta)(1+j \eta_+)}{1+(j+1)\eta_+}, \nonumber \\
& & \qquad \eta_{\pm} = \frac{(\alpha-\beta) - (\alpha \pm \gamma)}{(M+1)(\alpha \pm \gamma) - M(\alpha-\beta)}, \
\eeqn
where the initial conditions in (\ref{FUN_N_HALF_M1_2}) have been used \cite{REMARK}.

We can now proceed to express explicitly the intra- and inter-species lowest-order densities.
Because the 
auxiliary function $F_M$, together with the initial conditions $\tilde C_M=-\gamma$ and $D_M=\gamma$,
is proportional to the $N$-particle density (\ref{MB_DENS}),
the function $F_1$ is proportional to the intra-species two-body density (\ref{DENS_AB}).
Thus we readily have
\beqn\label{TWO_B_DENSITY}
& & \rho_{AB}(\x,\y) = M^2\left[\frac{(\alpha+C_1)^2-(D_1-C_1)^2}{\pi^2}\right]^\frac{3}{2}
e^{-(\alpha+C_1) (\x^2+\y^2) 
+2(D_1-C_1)\x \cdot \y}, \nonumber \\ 
& & = M^2\left[\frac{(\alpha+C_1)^2-(D_1-C_1)^2}{\pi^2}\right]^\frac{3}{2}
e^{- \frac{1}{2}[(\alpha+C_1)+(D_1-C_1)](\x-\y)^2}  e^{- \frac{1}{2}[(\alpha+C_1)-(D_1-C_1)](\x+\y)^2}, \
\eeqn
for the two-body density
and, upon one additional integration, 
\beqn\label{ONE_B_DENSITIES}
& & \qquad \rho_A(\x) = M 
\left[\frac{(\alpha+D_1)(\alpha+\tilde C_1)}{\pi(\alpha+C_1)}\right]^\frac{3}{2}
e^{-\frac{(\alpha+D_1)(\alpha+\tilde C_1)}{(\alpha+C_1)}\x^2}, \nonumber \\
& & \qquad \rho_B(\y) = M
\left[\frac{(\alpha+D_1)(\alpha+\tilde C_1)}{\pi(\alpha+C_1)}\right]^\frac{3}{2}
e^{-\frac{(\alpha+D_1)(\alpha+\tilde C_1)}{(\alpha+C_1)}\y^2}, \
\eeqn
for the one-body densities. 
We see 
that the two-body density can be viewed as an ellipsoid in the $\x$--$\y$ coordinates,
see the second line of (\ref{TWO_B_DENSITY}), 
whereas the one-body densities are isotropic Gaussians in $\x$ and $\y$ coordinates.

To complete the computation of the
densities we are left to determine $\tilde C_1$ and $D_1$ as a function of 
the mixture's frequencies $\Omega_{rel}$, $\Omega_{AB}$, and $\omega$ 
and the number of particles in each species $M$.
Using (\ref{WAVE_FUN_3}) we find
\beqn\label{ABC_PAR_1}
& & \alpha-\beta=\Omega_{rel}, \qquad
\alpha+\gamma = \frac{(M-1)\Omega_{rel}+\Omega_{AB}}{M}, \qquad 
\alpha-\gamma = \frac{(M-1)\Omega_{rel}+\omega}{M} \qquad 
\Longrightarrow \nonumber \\
& &  \qquad \eta_+ = \frac{\Omega_{rel}-\Omega_{AB}}{(M+1)\Omega_{AB}-\Omega_{rel}}, \qquad
 \eta_- = \frac{\Omega_{rel}-\omega}{(M+1)\omega-\Omega_{rel}}. \
\eeqn
From which we 
obtain the ingredients
\beqn\label{ABC_PAR_2}
& & \alpha+\tilde C_1 = \Omega_{rel} \, \frac{M\omega}{(M-1)\omega+\Omega_{rel}}, \nonumber \\
& & \alpha+D_1 = \Omega_{rel} \, \frac{M\Omega_{AB}}{(M-1)\Omega_{AB}+\Omega_{rel}}, \nonumber \\
& & \alpha+C_1 =  \frac{(\alpha+\tilde C_1) + (\alpha+D_1)}{2} =
\Omega_{rel} \, \frac{M[2(M-1)\omega\Omega_{AB}+(\omega+\Omega_{AB})\Omega_{rel}]}
{2[(M-1)\omega+\Omega_{rel}][(M-1)\Omega_{AB}+\Omega_{rel}]}, \nonumber \\
\eeqn
and combinations thereof
\beqn\label{ABC_PAR_3}
& & 2(D_1-C_1) = (\alpha + D_1) - (\alpha + \tilde C_1) = 
 \frac{M\Omega^2_{rel}(\Omega_{AB}-\omega)}
{[(M-1)\omega+\Omega_{rel}][(M-1)\Omega_{AB}+\Omega_{rel}]}, \nonumber \\
& & (\alpha+C_1)^2-(D_1-C_1)^2 = (\alpha+D_1)(\alpha+\tilde C_1) = 
\Omega^2_{rel} \, \frac{M^2\omega\Omega_{AB}}
{[(M-1)\omega+\Omega_{rel}][(M-1)\Omega_{AB}+\Omega_{rel}]}, \nonumber \\
& & \frac{(\alpha+D_1)(\alpha+\tilde C_1)}{(\alpha+C_1)} = 
\frac{2}{\frac{1}{\alpha+D_1}+\frac{1}{\alpha+\tilde C_1}} =
\Omega_{rel} \, \frac{2M\omega\Omega_{AB}}{2(M-1)\omega\Omega_{AB}+\Omega_{rel}(\omega+\Omega_{AB})} \
\eeqn
entering the expressions (\ref{TWO_B_DENSITY}) and (\ref{ONE_B_DENSITIES}).
We have now computed explicitly and analytically the lowest-order 
densities of the mixture.
 
With analytical expressions for the densities
one can examine various situations.
We wish to elaborate on two.
The two-body density couples the $A$ and $B$ species as soon as the inter-spices
interaction, $\lambda_2$, is present, see (\ref{TWO_B_DENSITY}).
This is because $\lambda_2\ne 0$ leads to $\Omega_{AB}\ne \omega$, see (\ref{MIX_FREQ}),
which implies $(D_1-C_1)\ne 0$, 
see the first line of (\ref{ABC_PAR_3}).
Furthermore, from the aspect ratio of the ellipsoid in the $\x$--$\y$ coordinates,
$(D_1-C_1)>0$ means that distinct particles tend to be together, i.e., inter-species attraction,
whereas $(D_1-C_1)<0$ means that distinct particles tend to be 
apart from each other, i.e., inter-species repulsion.
This is compatible with
(\ref{ABC_PAR_3}) which shows that $(D_1-C_1)$ is positive (negative)
when $\Omega_{AB}$ is bigger (smaller) than $\omega$, i.e., when $\lambda_2$ is positive (negative).
To remind, $\lambda_2>0$ signifies inter-species attraction 
and $\lambda_2<0$ repulsion. 

The second issue we would like to discuss is the so-called infinite-particle limit.
As mentioned in Sec.~\ref{INTRO},
this question has drawn much attention for single-species
Bose-Einstein condensates.
With the help of the present analytical many-body results,
and with the mean-field
solution of the subsequent section, 
we can, to the best of our knowledge for the first time,
give some concrete answers on what happens in the infinite-particle limit in a mixture.
Thus, holding the intra- and inter-species interaction parameters $\Lambda_1$ and $\Lambda_2$ fixed,
we have for the mixtue's frequencies in 
the infinite-particle limit
$\lim_{M \to \infty} \Omega_{rel} = \sqrt{\omega^2+2(\Lambda_1+\Lambda_2)}$
and 
$\lim_{M \to \infty} \Omega_{AB} = \sqrt{\omega^2+4\Lambda_2}$. 
Now, the ingredients (\ref{ABC_PAR_2}) 
entering the densities satisfy in the limit of an infinite number of particles
$\lim_{M \to \infty} (\alpha+C_1) = 
\lim_{M \to \infty} (\alpha+D_1) = 
\lim_{M \to \infty} (\alpha+\widetilde C_1) = \sqrt{\omega^2+2(\Lambda_1+\Lambda_2)}$,
and similarly in (\ref{ABC_PAR_3}).
Consequently, in the infinite-particle limit the expressions for the densities, 
see (\ref{DENS_A_B}) and (\ref{DENS_AB}),
depend on the interaction parameters only and simplify
\beqn\label{1B_DENS_LIM}
& &  \lim_{M \to \infty} \frac{\rho_A(\x)}{M} = 
\left(\frac{\sqrt{\omega^2 + 2(\Lambda_1+\Lambda_2)}}{\pi}\right)^{\frac{3}{2}} 
e^{-\sqrt{\omega^2 + 2(\Lambda_1+\Lambda_2)} \x^2}, \nonumber \\
& &  \lim_{M \to \infty} \frac{\rho_B(\y)}{M} = 
\left(\frac{\sqrt{\omega^2 + 2(\Lambda_1+\Lambda_2)}}{\pi}\right)^{\frac{3}{2}}
e^{-\sqrt{\omega^2 + 2(\Lambda_1+\Lambda_2)} \y^2},  \
\eeqn
and
\beq\label{1B_1B_DENS_LIM}
 \lim_{M \to \infty} \frac{\rho_{AB}(\x,\y)}{M^2} = 
\left(\frac{\sqrt{\omega^2 + 2(\Lambda_1+\Lambda_2)}}{\pi}\right)^3 
e^{-\sqrt{\omega^2 + 2(\Lambda_1+\Lambda_2)} (\x^2+\y^2)}.
\eeq
In particular, because $\lim_{M \to \infty} (D_1-C_1) = 0$
the inter-species two-body density is seen to factorize to a product
of two one-body densities,
independent of the magnitude of the inter-species interaction parameter $\Lambda_2$.
We would come back to this point in the following section,
when the mean-field solution of the mixture's HIM model is to be derived and analyzed.

\subsection{Mean-field (Gross-Pitaevskii) solution}

At the other end of the exact, many-body solution of the mixture's HIM model lies
the Gross-Pitaevskii, mean-field solution.
In the mean-field theory the many-particle wave-function is approximated as a product state,
where all the bosons of the $A$ species lies in the one and the same orbital,
and all the bosons of the $B$ species similarly lie in one orbital
which is generally different than the $A$ species one.
In our case of a symmetric mixture (\ref{HAM_MIX}),
excluding a solution where demixing of the two species occurs in the ground state,
the mean-field ansatz for the mixture is the 
product wave-function where, 
due to symmetry between the $A$ and $B$ species, 
each of the 
species occupies the same spatial function 
\beq\label{MIX_WAV_GP}
 \Phi^{GP}(\x_1,\ldots,\x_M,\y_1,\ldots,\y_M) = 
\prod_{j=1}^M \phi(\x_j) \prod_{k=1}^M \phi(\y_k). 
\eeq
The Gross-Pitaevskii energy functional of the mixture thus simplifies and reads
\beqn\label{MIX_EF_GP_1}
& & \varepsilon^{GP} =
\frac{1}{2}\Bigg[\int d\x \phi^\ast(\x) \left(-\frac{1}{2} \frac{\partial^2}{\partial \x^2} 
+ \frac{1}{2} \omega^2 \x^2 \right) \phi(\x) + \frac{\Lambda_1}{2}
	\int d\x d\x' |\phi(\x)|^2 |\phi(\x')|^2 (\x-\x')^2 + \nonumber \\
& & + \int d\y \phi^\ast(\y) \left(-\frac{1}{2} \frac{\partial^2}{\partial \y^2} 
+ \frac{1}{2} \omega^2 \y^2 \right) \phi(\y) + \frac{\Lambda_1}{2}
	\int d\y d\y' |\phi(\y)|^2 |\phi(\y')|^2 (\y-\y')^2\Bigg] + \nonumber \\
& & \qquad \qquad + \frac{\Lambda_2}{2}
	\int d\x d\y |\phi(\x)|^2 |\phi(\y)|^2 (\x-\y)^2 = \nonumber \\
& & = \int d\r \phi^\ast(\r) \left(-\frac{1}{2} \frac{\partial^2}{\partial \r^2}
+ \frac{1}{2} \omega^2 \r^2 \right) \phi(\r) + \frac{1}{2}(\Lambda_1 + \Lambda_2)
	\int d\r d\r' |\phi(\r)|^2 |\phi(\r')|^2 (\r-\r')^2,
\eeqn 
where $\r$ represents the coordinate of any of the particles.
To remind, $\varepsilon^{GP}$ is the total mean-field energy of the mixture divided by the 
number of particle $N=2M$.
Consequently and side by side, 
the two coupled Gross-Pitaevskii equations of the mixture degenerate to one
\beqn\label{MIX_EQ_GP_1}
 \left\{ -\frac{1}{2} \frac{\partial^2}{\partial \r^2} + \frac{1}{2} \omega^2 \r^2 + 
(\Lambda_1 + \Lambda_2) \int d\r' |\phi(\r')|^2 (\r-\r')^2 \right\} \phi(\r) =
\mu \phi(\r), 
\eeqn
where $\mu$ is the chemical potential of each species.
The solution of 
(\ref{MIX_EQ_GP_1}) follows exactly the same way as for the HIM problem \cite{HIM_Cohen},
and is now briefly followed for completeness.
Expanding the interaction term in (\ref{MIX_EQ_GP_1}) we find
\beq\label{MIX_EQ_GP_2}
\left\{ -\frac{1}{2} \frac{\partial^2}{\partial \r^2} 
+ \frac{1}{2} \left[\omega^2 + 2(\Lambda_1+\Lambda_2)\right]\r^2 \right\} \phi(\r) =
\left[\mu - (\Lambda_1+\Lambda_2) \int d\r' |\phi(\r')|^2 \r'^2 \right] \phi(\r). 
\eeq
The solution of (\ref{MIX_EQ_GP_2}) 
is the Gaussian function.
\beqn\label{MIX_GP_OR}
& & \phi(\r) = \left(\frac{\sqrt{\omega^2 + 2(\Lambda_1+\Lambda_2)}}{\pi}\right)^{\frac{3}{4}}
e^{-\frac{1}{2}\sqrt{\omega^2 + 2(\Lambda_1+\Lambda_2)} \r^2} = \nonumber \\
& & \qquad = \left(\frac{\sqrt{\Omega_{rel}^2 - 2\lambda_1}}{\pi}\right)^{\frac{3}{4}}
e^{-\frac{1}{2}\sqrt{\Omega_{rel}^2 - 2\lambda_1} \r^2}. \
\eeqn
Since the orbital (\ref{MIX_GP_OR}) is an even function,
there is no linear in $\r$ 
term in (\ref{MIX_EQ_GP_2}).
We can now evaluate the integral $\int d\r' |\phi(\r')|^2 \r'^2 = \frac{3}{2\sqrt{\omega^2 + 2(\Lambda_1+\Lambda_2)}}$ in (\ref{MIX_EQ_GP_2}) and determine the chemical potential
\beq\label{MIX_GP_MU}
 \mu = \frac{3}{2}\sqrt{\omega^2 + 2(\Lambda_1+\Lambda_2)} + \frac{3}{4}
\frac{\Lambda_1+\Lambda_2}{\sqrt{\omega^2 + 2(\Lambda_1+\Lambda_2)}}
= \frac{3}{4} 
\frac{2\omega^2 + 3(\Lambda_1+\Lambda_2)}{\sqrt{\omega^2 + 2(\Lambda_1+\Lambda_2)}}. 
\eeq 
Indeed, for $\lambda_1=\lambda_2$ the single-species HIM chemical potential \cite{HIM_Cohen} is found.
With this, the mean-field energy is given by
\beqn\label{MIX_E_GP}
 & & \varepsilon^{GP} = 
\left\{\mu - \frac{1}{2}(\Lambda_1+\Lambda_2)\int d\r d\r' |\phi(\r)|^2 |\phi(\r')|^2 (\r-\r')^2 \right\} = \nonumber \\
& & \qquad = 
\frac{3}{2}\sqrt{\omega^2 + 2(\Lambda_1+\Lambda_2)} = \frac{3}{2}\sqrt{\Omega_{rel}^2 - 2\lambda_1},
\eeqn
where $\int d\r d\r' |\phi(\r)|^2 |\phi(\r')|^2 (\r-\r')^2 = \frac{3}{\sqrt{\omega^2 + 2(\Lambda_1+\Lambda_2)}}$ is used. 
In the specific case where $\Lambda_1+\Lambda_2=0$,  
the mean-field energy becomes that of the non-interacting system.
This reminds the properties of the many-body energy discussed above,
see (\ref{HIM_MIX_GS_E0}) and 
(\ref{HIM_MIX_GS_E0_INF}) and associated text. 

We can now compare the exact, many-body energy $E$ and 
the mean-field energy per particle $\varepsilon^{GP}$.
Of course, the many-body energy is always lower, for repulsive as well as for attractive particle-particle interactions,
than the mean-field energy because of the variational principle.
In the infinite-particle limit we find when $\Lambda_1$ and $\Lambda_2$ are held constant that 
\beq\label{MIX_E_GP_INF}
 \lim_{N\to \infty} \frac{E}{N} = \varepsilon^{GP},
\eeq
which establishes the connection between the exact energy and mean-field (Gross-Pitaevskii) 
energy per particle in this limit for the mixture.

We now discuss the one-body density.
From (\ref{MIX_GP_OR}) we have
\beqn\label{MIX_GP_DENS}
& & \rho^{GP}(\r) = |\phi^{GP}(\r)|^2 = \left(\frac{\sqrt{\omega^2 + 2(\Lambda_1+\Lambda_2)}}{\pi}\right)^{\frac{3}{2}}
e^{-\sqrt{\omega^2 + 2(\Lambda_1+\Lambda_2)} \r^2} = \nonumber \\
& & \qquad = \left(\frac{\sqrt{\Omega_{rel}^2 - 2\lambda_1}}{\pi}\right)^{\frac{3}{2}}
e^{-\sqrt{\Omega_{rel}^2 - 2\lambda_1} \r^2}. \
\eeqn
From the mean-field solution we see that the density narrows for overall attractive interactions and broadens for repulsive ones.
To be more precise, the density narrows for 
$\Lambda_1+\Lambda_2>0$ 
and broadens for 
$\Lambda_1+\Lambda_2<0$.
Within the mean-field theory this is the manifestation of the self-consistency 
used to solve the 
non-linear Schr\"odinger (Gross-Pitaevskii) equation
and find the
orbital (\ref{MIX_GP_OR}).
Furthermore, 
from (\ref{MIX_GP_DENS}) and using (\ref{MIX_WAV_GP}) we can 
write for the lowest-order densities of the mixture 
within the mean-field theory,
$\rho_A^{GP}(\x) = M\rho^{GP}(\x)$,
$\rho_B^{GP}(\y) = M\rho^{GP}(\y)$,
and
$\rho_{AB}^{GP}(\x,\y) = M^2\rho^{GP}(\x)\rho^{GP}(\y)$.
As expected, 
for mixtures with a finite number of particles, 
the mean-field densities 
differ from their many-body counterparts (\ref{ONE_B_DENSITIES}) and (\ref{TWO_B_DENSITY}).
For instance, in contrast to 
the two-body inter-species density $\rho_{AB}(\x,\y)$,
the mean-field density $\rho_{AB}^{GP}(\x,\y)$ is always factorized to
a product of two one-body densities. 
Finally, in the limit of an infinite number of particles the many-body 
(\ref{1B_1B_DENS_LIM}) and (\ref{1B_DENS_LIM}) and the Gross-Pitaevskii densities
coincide in the sense that what 
one might have suspected
holds true, namely
\beq\label{1B_DENS_LIM_GP}
 \lim_{M \to \infty} \frac{\rho_A(\x)}{M} = \rho^{GP}(\x), \qquad
 \lim_{M \to \infty} \frac{\rho_B(\y)}{M} = \rho^{GP}(\y), 
\eeq
and
\beq\label{1B_1B_DENS_LIM_GP}
 \lim_{M \to \infty} \frac{\rho_{AB}(\x,\y)}{M^2} = \rho^{GP}(\x)\rho^{GP}(\y).
\eeq
With (\ref{MIX_E_GP_INF}), (\ref{1B_DENS_LIM_GP}), and (\ref{1B_1B_DENS_LIM_GP}),
we have proved and established 
that the many-body energy and densities 
of the mixture
coincide with their mean-field values
in the infinite-particle limit.
This constitutes a generalization for mixtures of Bose-Einstein condensates, 
at least within the mixture's HIM model,
of what is known in the literature for single-species 
trapped Bose-Einstein condensates \cite{Yngvason_PRA,Lieb_PRL}.

\subsection{Center-of-mass variance and uncertainty product}\label{CENT_MIX_VAR}

So far we have discussed the energy and densities of the mixture,
and have established the relations in the infinite-particle limit 
in which the exact solution and the Gross-Pitaevskii theory coincide.
As has recently been shown for single-species Bose-Einstein condensates,
the variance of many-particle operators can deviate strongly
when the many-body and mean-field treatments are compared \cite{Variance,TD_Variance}.
These quantities have recently been found to be sensitive to correlations in
single-species Bose-Einstein condensates even 
in the infinite-particle limit. 
The many-body and mean-field analytical solutions of the 
mixture allow 
us to examine directly the variances of the center-of-mass position and momentum operators of the mixture,
as well as their uncertainty product.
We shall keep the discussion concise.

Consider the center-of-mass position and momentum operators of the mixture
\beqn\label{CM_MIX_X_P}
 & & {\hat \R}_{CM} = 
\frac{1}{N} \sum_{j=1}^M (\x_j+\y_j) = \frac{1}{\sqrt{N}} \Q_N, \quad
 {\hat \P}_{CM} = 
\sum_{j=1}^M \left(\frac{1}{i}\frac{\partial}{\partial \x_j}+\frac{1}{i}\frac{\partial}{\partial \y_j}\right) =
\sqrt{N} \frac{1}{i}\frac{\partial}{\partial \Q_N}, \nonumber \\
& & \qquad \bl{\hat \R}_{CM}, {\hat \P}_{CM}\br = \bi, \quad \forall N, \
\eeqn
where $\bi$ (and analogously $\bo$ below) is 
a shorthand symbol for $i$ in each of the 
three Cartesian components.
${\hat \R}_{CM}$ and ${\hat \P}_{CM}$ are one-particle operators 
in the sense that they are linear with respect to all particles in the mixture.
To compute their variances we 
need also 
the operators squares, ${\hat \R}_{CM}^2$ and ${\hat \P}_{CM}^2$,
which are two-particle operators.
These would 
lead to differences between the many-body and mean-field results;
see for an extended discussion of the matter in the case 
of single-species Bose-Einstein condensates \cite{Variance,TD_Variance}.

We can now compute the variances of the center-of-mass position and momentum operators,
\beqn\label{CM_VAR_MB}
& & \Delta^2_{\hat \R_{CM}} = \langle\Psi|\hat \R_{CM}^2|\Psi\rangle -
 \langle\Psi|\hat \R_{CM}|\Psi\rangle^2 = \frac{1}{N} \times \frac{1}{2\omega} \bo, \nonumber \\
& & \Delta^2_{\hat \P_{CM}} = \langle\Psi|\hat \P_{CM}^2|\Psi\rangle -
 \langle\Psi|\hat \P_{CM}|\Psi\rangle^2 = N \times \frac{\omega}{2} \bo, \quad \forall N, \
\eeqn
which are straightforwardly obtained due to the separability of the wave-function (\ref{WAVE_FUN_1}).
For comparison, 
the corresponding expressions at the Gross-Pitaevskii level are computed from 
the mean-field wave-function (\ref{MIX_WAV_GP}) and given by
\beqn\label{CM_VAR_GP}
& & \!\!\!\!\!\!\!\! \Delta^2_{\hat \R_{CM},GP} = \langle\Phi^{GP}|\hat \R_{CM}^2|\Phi^{GP}\rangle -
 \langle\Phi^{GP}|\hat \R_{CM}|\Phi^{GP}\rangle^2 = \frac{1}{N} \times \frac{1}{2\sqrt{\omega^2+2(\Lambda_1+\Lambda_2)}} \bo, \nonumber \\
& & \!\!\!\!\!\!\!\! \Delta^2_{\hat \P_{CM},GP} = \langle\Phi^{GP}|\hat \P_{CM}^2|\Phi^{GP}\rangle -
 \langle\Phi^{GP}|\hat \P_{CM}|\Phi^{GP}\rangle^2 = N \times \frac{\sqrt{\omega^2+2(\Lambda_1+\Lambda_2)}}{2} \bo, \quad \forall N. \
\eeqn
Comparing (\ref{CM_VAR_MB}) and (\ref{CM_VAR_GP}) 
we see how the self-consistency of the Gross-Pitaevskii orbtial (\ref{MIX_GP_OR}),
whose shape depends on the sum of the intra- and inter-species 
interaction parameters $\Lambda_1+\Lambda_2$,
alters the values of the mean-field 
variances in comparison with the exact result.
This is an interesting result for the mixture
which is 
best expressed by the respective ratios
\beq\label{CM_VAR_MB_VS_GP}
\frac{\Delta^2_{\hat \R_{CM},GP}}{\Delta^2_{\hat \R_{CM}}} = 
\frac{1}{\sqrt{1+\frac{2}{\omega^2}(\Lambda_1+\Lambda_2)}}\bo, \qquad
\frac{\Delta^2_{\hat \P_{CM},GP}}{\Delta^2_{\hat \P_{CM}}} =
\sqrt{1+\frac{2}{\omega^2}(\Lambda_1+\Lambda_2)}\bo, \quad \forall N,
\eeq
in which the total number of particles does not appear.
In particular for predominantly attractive mixtures, 
$\Lambda_1+\Lambda_2 > 0$,
since the sum $\Lambda_1+\Lambda_2$ is unbound from above
the ratio between 
the mean-field and exact
center-of-mass position variances 
(\ref{CM_VAR_MB_VS_GP}) can be as small as one wishes.
However,
the ratio of the center-of-mass momentum variances
(\ref{CM_VAR_MB_VS_GP})
is then, inevitably, very large. 

We can now discuss the uncertainty product.
Here, because the trap and all interactions are harmonic,
the center-of-mass wave-function is a Gaussian and
the mean-field orbital is a Gaussian as well,
although being dressed by the particle-particle interactions.
Consequently,
\beq\label{UP_MB_GP}
\Delta^2_{\hat \R_{CM}}\Delta^2_{\hat \P_{CM}} = \frac{1}{4}\bo, \qquad 
\Delta^2_{\hat \R_{CM},GP}\Delta^2_{\hat \P_{CM},GP} = \frac{1}{4}\bo, \quad \forall N,
\eeq
irrespective to the interactions between the particles.
The comparison between the exact and mean-field solutions
in terms of the uncertainty product might erroneously imply
that there are no correlations in the mixture,
especially in the limit of an infinite number of particles.
Here, the correlations which survive the infinite-particle limit can for instance be seen in 
the ratios of the variances 
(\ref{CM_VAR_MB_VS_GP}).
Although not discussed in \cite{HIM_Cohen},
this generalizes the situation for the single-species HIM 
to the mixture's HIM model.

A final note.
The result (\ref{UP_MB_GP}) raises the useful question what happens for general intra- and inter-species interactions.
Is this equivalence between the center-of-mass uncertainty products
at the exact and mean-field levels lifted?
The answer goes beyond the solution of the HIM model for mixtures but, 
together with the related issue of an explicit construction of the center-of-mass separability in the generic case,
is presented for completeness in the Appendix. 

\section{Concluding Remarks}\label{CON_REM}

We have presented in this work a solvable quantum model for 
a mixture of two trapped Bose-Einstein condensates.
The model consists of 
two kinds of identical bosons, $A$ and $B$, 
held in a harmonic potential and interacting 
by harmonic particle-particle interactions,
namely, the harmonic-interaction model for mixtures.
We have concentrated in 
the present investigation
on the case of a symmetric mixture
in which the number of bosons and interactions within species $A$ 
are the same as for species $B$.
By generalizing the treatment of Cohen and Lee \cite{HIM_Cohen} to mixtures,
closed form expressions for the ground-state energy, wave-function, 
and lowest-order intra- and inter-species densities are obtained.
These quantities are analyzed analytically as a function of the number of particles and the
intra- and inter-species interactions.

Aside from the many-body solution,
we have also obtained analytically the Gross-Pitaevskii solution for the ground-state,
as far as demixing of the two Bose-Einstein condensates is excluded.
This has allowed us to compare the exact, many-body and mean-field solutions
for any number of particles.
In particular, keeping the products of the number of particles times the intra-species and inter-species
interaction strengths fixed, while increasing the number of particles in the mixture,
we were able to prove that the exact ground-state energy and 
lowest-order intra-species and inter-species densities converge 
at the infinite-particle limit to the results of the Gross-Pitaevskii theory for the mixture. 
This holds as a particular generalization for bosonic mixtures 
of the known literature results for single-species bosons \cite{Yngvason_PRA,Lieb_PRL}. 
On the other end, the separability of the mixture's center-of-mass coordinate 
is used to show that the result for the variance of many-particle operators in the mixture
obtained within Gross-Pitaevskii theory for mixtures can deviate substantially
from the exact, many-body result, even in the limit of an infinite number of particles. 
The present analytical results hence show that many-body correlations exist in the ground state 
of a trapped mixture of Bose-Einstein condensates made of any number of particles,
thus generalizing our recent result for single-species bosons \cite{Variance}.

As an outlook we mention the asymmetric mixture, out-of-equilibrium dynamics, and more,
all within the harmonic-interaction model for mixtures.
In particular, an impurity made of one or two particles 
would be interesting to solve,
once the techniques used above are adapted to asymmetric integrations.
Last but not least,
we hope that the present analytical results and proofs
obtained in the limit of an infinite number of particles 
for a particular solvable model would stimulate generalizations 
for mixtures with short-range interactions,
in as much as single-species bosons 
were offered 
mathematically rigorous results in this limit in the literature.

\section*{Acknowledgements}

This research was supported by the Israel Science Foundation (Grant No. 600/15). 
Partial financial support by the Deutsche Forschungsgemeinschaft (DFG) is acknowledged.
O.E.A. is grateful to the continuous
hospitality of the Lewiner Institute for Theoretical Physics
(LITP) at the Department of Physics, Technion.

\appendix

\section{Center-of-mass separability and 
uncertainty product for a mixture with generic particle-particle interactions}\label{CENT_GER_UP}

Let the Hamiltonian of the mixture with generic particle-particle 
interactions be
\beqn\label{HAM_MIX_G}
& & \hat H(\x_1,\ldots,\x_M,\y_1,\ldots,\y_M) = 
 \sum_{j=1}^M 
\left[\left( -\frac{1}{2} \frac{\partial^2}{\partial \x_j^2} + \frac{1}{2} \omega^2 \x_j^2 \right) +
\left( -\frac{1}{2} \frac{\partial^2}{\partial \y_j^2} + \frac{1}{2} \omega^2 \y_j^2 \right)\right] + \nonumber \\
& & \qquad + \lambda_1 \sum_{1 \le j < k}^M \left[ W_1(\x_j-\x_k) + W_1(\y_j-\y_k) \right]  
+ \lambda_2 \sum_{j=1}^M \sum_{k=1}^M  W_2(\x_j-\y_k).
\eeqn
With the ansatz $\Phi^{GP}(\x_1,\ldots,\x_M,\y_1,\ldots,\y_M) = 
\prod_{j=1}^M \phi(\x_j) \prod_{k=1}^M \phi(\y_k)$
[Eq.~(\ref{MIX_WAV_GP})], i.e., when excluding demixing in the 
ground state,
the mixture's 
Gross-Pitaevskii energy functional reduces to
\beqn\label{MIX_EF_GP_W}
& & \varepsilon^{GP} = \int d\r \phi^\ast(\r) \left(-\frac{1}{2} \frac{\partial^2}{\partial \r^2}
+ \frac{1}{2} \omega^2 \r^2 \right) \phi(\r) + \nonumber \\ 
& & \qquad + \frac{1}{2} \int d\r d\r' |\phi(\r)|^2 |\phi(\r')|^2
\left[\Lambda_1 W_1(\r-\r') + \Lambda_2 W_2(\r-\r') \right], \
\eeqn 
and the resulting Gross-Pitaevskii equations degenerate to one
\beq\label{MIX_EQ_GP_G_1}
 \left\{ -\frac{1}{2} \frac{\partial^2}{\partial \r^2} + \frac{1}{2} \omega^2 \r^2 + 
\int d\r' |\phi(\r')|^2
\left[\Lambda_1 W_1(\r-\r')
+ \Lambda_2 W_2(\r-\r') \right] \right\} \phi(\r) = \mu \phi(\r). \ 
\eeq
The generic intra-species $W_1$ and inter-species $W_2$ interactions
are seen to dress the Gross-Pitaevskii orbital $\phi$ of the $A$ and $B$ species in the mixture 
to a shape different than the Gaussian in (\ref{MIX_GP_OR}).
The reason is that the mean-field potential 
$\int d\r' |\phi(\r')|^2 [\Lambda_1 W_1(\r-\r') + \Lambda_2 W_2(\r-\r')]$
for generic interactions is different than $\r^2$.

Next, we comment on the separability of the center-of-mass coordinate $\Q_N$ in the ground state
of the Hamiltonian (\ref{HAM_MIX_G}),
namely, 
we show that the coordinates needed to express the interaction terms are 
all but $\Q_N$.
This has been shown explicitly in the main text by diagonalizing the interaction terms of the
mixture's HIM, but such diagonalizing cannot be made with generic particle-particle interactions.
 
This is clearly the case for the intra-species interaction terms $W_1(\x_j-\x_k)$ and $W_1(\y_j-\y_k)$
on the account of the relative coordinates
$\Q_1,\ldots,\Q_{M-1}$ and $\Q_M,\ldots,\Q_{2(M-1)}$
of each species separately, see (\ref{MIX_COOR}).
But what about the inter-species interaction terms $W_2(\x_j-\y_k)$?
To express them, adding $\Q_{N-1}$, the relative coordinate 
between the center-of-mass of the $A$ and the center-of-mass of the $B$ species, suffices.
To illustrate this point, 
consider the 
inter-species relative coordinate $\x_M-\y_M$ which can be expressed as follows:
\beq\label{X1_Y1_TERM}
 \x_M-\y_M = \sqrt{\frac{M-1}{M}}[\Q_{M-1}-\Q_{2(M-1)}] + \sqrt{\frac{2}{M}} \Q_{N-1}. 
\eeq
The other inter-species relative coordinate $\x_j-\y_k$ can be expressed analogously,
using the fact that the relative coordinates of each species separately,
$\Q_1,\ldots,\Q_{M-1}$ and $\Q_M,\ldots,\Q_{2(M-1)}$,
can be intermixed at will
to generate any of the permutations of $\Q_{M-1}$ and of $\Q_{2(M-1)}$
with respect to the $\x_1,\ldots,\x_M$ and $\y_1,\ldots,\y_M$ coordinates, respectively.

All in all, the Hamiltonian (\ref{HAM_MIX_G}) is separable in the sense that
$\hat H = \left(-\frac{1}{2}\frac{\partial^2}{\partial \Q^2} 
+ \frac{1}{2}\omega^2 \Q_N^2\right) + \hat H_{rel}(\Q_1,\ldots,\Q_{N-1})$,
and so does its ground state, 
$\Psi = \left(\frac{\omega}{\pi}\right)^{\frac{3}{4}} 
e^{-\frac{1}{2}\omega\Q_N^2} \Psi_{rel}(\Q_1,\ldots,\Q_{N-1})$.
Here `{\it rel}' stands 
for all but center-of-mass quantities.
Consequently,
for mixtures with generic particle-particle interactions $W_1$ and $W_2$
the uncertainty products computed at the many-body and mean-field levels, 
not just the individual center-of-mass position and momentum quantities,
like the ratios (\ref{CM_VAR_MB_VS_GP}) discussed above, 
differ from each other even in the infinite-particle limit.
Put in formulas,
\beq\label{UP_MB_GP_GEN}
\Delta^2_{\hat \R_{CM},GP}\Delta^2_{\hat \P_{CM},GP} > 
\Delta^2_{\hat \R_{CM}}\Delta^2_{\hat \P_{CM}} = \frac{1}{4}\bo, \quad \forall N,
\eeq
which concludes our 
discussion.


\begin{thebibliography}{99}

\bibitem{Be1} C. J. Myatt, E. A. Burt, R. W. Ghrist, E. A. Cornell, and C. E. Wieman,
                   {\rm Production of Two Overlapping Bose-Einstein Condensates by Sympathetic Cooling},
                   Phys. Rev. Lett. {\bf 78}, 586 (1997).

\bibitem{Be2} D. M. Stamper-Kurn, M. R. Andrews, A. P. Chikkatur, S. Inouye, 
                   H.-J. Miesner, J. Stenger, and W. Ketterle,
                   {\rm Optical Confinement of a Bose-Einstein Condensate},
                   Phys. Rev. Lett. {\bf 80}, 2027 (1998).

\bibitem{Be3} D. S. Hall, M. R. Matthews, J. R. Ensher, C. E. Wieman, and E. A. Cornell,
                   {\rm Dynamics of Component Separation in a Binary Mixture of Bose-Einstein Condensates},
                   Phys. Rev. Lett. {\bf 81}, 1539 (1998).

\bibitem{Be4} M. R. Matthews, B. P. Anderson, 
                   P. C. Haljan, D. S. Hall, C. E. Wieman, and E. A. Cornell,
                   {\rm Vortices in a Bose-Einstein Condensate},
                   Phys. Rev. Lett. {\bf 83}, 2498 (1999).

\bibitem{Bt1} T.-L. Ho and V. B. Shenoy,
                    {\rm Binary Mixtures of Bose Condensates of Alkali Atoms},
                    Phys. Rev. Lett. {\bf 77}, 3276 (1996).

\bibitem{Bt2} B. D. Esry, C. H. Greene, J. P. Burke, Jr., and J. L. Bohn, 
                    {\rm Hartree-Fock Theory for Double Condensates},
                    Phys. Rev. Lett. {\bf 78}, 3594 (1997).

\bibitem{Bt3} Th. Busch, J. I. Cirac, V. M. P\'erez-Garc\'ia, and P. Zoller,
                   {\rm Stability and collective excitations of 
                   a two-component Bose-Einstein condensed gas: A moment approach},
                   Phys. Rev. A {\bf 56}, 2978 (1997).

\bibitem{Bt4} H. Pu and N. P. Bigelow, 
                    {\rm Properties of Two-Species Bose Condensates},
                    Phys. Rev. Lett. {\bf 80}, 1130 (1998).

\bibitem{Bt5} H. Pu and N. P. Bigelow, 
                    {\rm Collective Excitations, Metastability, and Nonlinear 
                    Response of a Trapped Two-Species Bose-Einstein Condensate},
                    Phys. Rev. Lett. {\bf 80}, 1134 (1998).

\bibitem{Bt6} E. Timmermans, 
                    {\rm Phase Separation of Bose-Einstein Condensates},
                    Phys. Rev. Lett. {\bf 81}, 5718 (1998).

\bibitem{Bt7} P. \"Ohberg and L. Santos, 
                   {\rm Dark Solitons in a Two-Component Bose-Einstein Condensate},
                   Phys. Rev. Lett. {\bf 86}, 2918 (2001).

\bibitem{Bt8} E. Altman, W. Hofstetter, E. Demler, and M. D. Lukin,
                    {\rm Phase diagram of two-component bosons on an optical lattice},
                    New J. Phys. {\bf 5}, 113 (2003).

\bibitem{Bt9} A. B. Kuklov and B. V. Svistunov,
                    {\rm Counterflow Superfluidity of Two-Species Ultracold 
                    Atoms in a Commensurate Optical Lattice},
                    Phys. Rev. Lett. {\bf 90}, 100401 (2003).

\bibitem{Bt10} K. Kasamatsu, M. Tsubota, and M. Ueda,
                     {\rm Vortex Phase Diagram in Rotating Two-Component Bose-Einstein Condensates},
                     Phys. Rev. Lett. {\bf 91}, 150406 (2003).

\bibitem{Bt11} H. T. Ng, C. K. Law, and P. T. Leung,
                     {\rm Quantum-correlated double-well tunneling of two-component Bose-Einstein condensates},
                     Phys. Rev. A {\bf 68}, 013604 (2003).

\bibitem{Bt12} A. Eckardt, C. Weiss, and M. Holthaus,
                     {\rm Ground-state energy and depletions for a dilute binary Bose gas},
                     Phys. Rev. A {\bf 70}, 043615 (2004).

\bibitem{Bt13} A. Kuklov, N. Prokof’ev, and B. Svistunov,
                     {\rm Commensurate Two-Component Bosons in an Optical Lattice: Ground State Phase Diagram},
                     Phys. Rev. Lett. {\bf 92}, 050402 (2004).

\bibitem{Bt14} A. Isacsson, M.-C. Cha, K. Sengupta, and S. M. Girvin,
                     {\rm Superfluid-insulator transitions of two-species bosons in an optical lattice},
                     Phys. Rev. B {\bf 72}, 184507 (2005).

\bibitem{Bt15} O. E. Alon, A. I. Streltsov, and L. S. Cederbaum,
                     {\rm Demixing of Bosonic Mixtures in Optical Lattices from Macroscopic to Microscopic},
                     Phys. Rev. Lett. {\bf 97}, 230403 (2006).

\bibitem{Bt16} A. Gubeskys, B. A. Malomed, and I. M. Merhasin,
                    {\rm Two-component gap solitons in two- and one-dimensional Bose-Einstein condensates},
                    Phys. Rev. A {\bf 73}, 023607 (2006).
 
\bibitem{Bt17} K. Kasamatsu and M. Tsubota,
                    {\rm Modulation instability and solitary-wave formation in two-component Bose-Einstein condensates},
                     Phys. Rev. A {\bf 74}, 013617 (2006).

\bibitem{Bt18} H. A. Cruz, V. A. Brazhnyi, V. V. Konotop, G. L. Alfimov, and M. Salerno,
                    {\rm Mixed-symmetry localized modes and breathers 
                    in binary mixtures of Bose-Einstein condensates in optical lattices},
                    Phys. Rev. A {\bf 76}, 013603 (2007).

\bibitem{Bt19} K. Nho and D. P. Landau,
                     {\rm Finite-temperature properties of binary mixtures of two Bose-Einstein condensates},
                     Phys. Rev. A {\bf 76}, 053610 (2007). 

\bibitem{Bt20} O. E. Alon, A. I. Streltsov, and L. S. Cederbaum,
                     {\rm Multiconfigurational time-dependent Hartree method for 
                     mixtures consisting of two types of identical particles},
                     Phys. Rev. A. {\bf 76}, 062501 (2007).

\bibitem{Bt21} L. S. Cederbaum, A. I. Streltsov, Y. B. Band, and O. E. Alon,
                     {\rm Interferences in the Density of Two Bose-Einstein Condensates Consisting of Identical or Different Atoms},
                     Phys. Rev. Lett. {\bf 98}, 110405 (2007).

\bibitem{Bt22} A. R. Sakhel, J. L. DuBois, and H. R. Glyde,
                     {\rm Condensate depletion in two-species Bose gases: A variational quantum Monte Carlo study},
                     Phys. Rev. A {\bf 77}, 043627 (2008).

\bibitem{Bt23} S. Z\"ollner, H.-D. Meyer, and P. Schmelcher,
                     {\rm Composite fermionization of one-dimensional Bose-Bose mixtures},
                     Phys. Rev. A {\bf 78}, 013629 (2008).

\bibitem{Bt24} B. Ole\'s and K. Sacha,
                     {\rm $N$-conserving Bogoliubov vacuum of a two-component 
                     Bose-Einstein condensate: density fluctuations close to a phase-separation condition},
                     J. Phys. A {\bf 41}, 145005 (2008).

\bibitem{Bt25} P. Buonsante, S. M. Giampaolo, F. Illuminati, V. Penna, and A. Vezzani,
                     {\rm Mixtures of Strongly Interacting Bosons in Optical Lattices},
                     Phys. Rev. Lett. {\bf 100}, 240402 (2008).

\bibitem{Bt26}  I. I. Satija, R. Balakrishnan, P. Naudus, J. Heward, M. Edwards, and C. W. Clark,
                      {\rm Symmetry-breaking and symmetry-restoring dynamics of a mixture 
                      of Bose-Einstein condensates in a double well},
                      Phys. Rev. A {\bf 79}, 033616 (2009).

\bibitem{Bt27} R. Navarro, R. Carretero-Gonz\'alez, and P. G. Kevrekidis,
                     {\rm Phase separation and dynamics of two-component Bose-Einstein condensates},
                     Phys. Rev. A {\bf 80}, 023613 (2009).

\bibitem{Bt28} Y. Hao and S. Chen,
                     {\rm Density-functional theory of two-component Bose gases in one-dimensional harmonic traps},
                     Phys. Rev. A {\bf 80}, 043608 (2009).

\bibitem{Bt30} M. D. Girardeau,
                     {\rm Pairing, Off-Diagonal Long-Range Order, and Quantum Phase Transition in Strongly 
                     Attracting Ultracold Bose Gas Mixtures in Tight Waveguides},
                     Phys. Rev. Lett. {\bf 102}, 245303 (2009).

\bibitem{Bt30h} J. Smyrnakis, S. Bargi, G. M. Kavoulakis, M. Magiropoulos, K. K\"arkk\"ainen, and S. M. Reimann,
                       {\rm Mixtures of Bose Gases Confined in a Ring Potential},
                       Phys. Rev. Lett. {\bf 103}, 100404 (2009).

\bibitem{Bt31} E. Tempfli, S. Z\"ollner, and P. Schmelcher,
                     {\rm Binding between two-component bosons in one dimension},
                     New J. Phys. {\bf 11}, 073015 (2009).

\bibitem{Bt32} A. C. Pflanzer, S. Z\"ollner, and P. Schmelcher,
                     {\rm Interspecies tunneling in one-dimensional Bose mixtures},
                     Phys. Rev. A {\bf 81}, 023612 (2010).

\bibitem{Bt33} M. D. Girardeau and G. E. Astrakharchik, 
                     {\rm Ground state of a mixture of two bosonic 
                     Calogero-Sutherland gases with strong odd-wave interspecies attraction},
                     Phys. Rev. A {\bf 81}, 043601 (2010).

\bibitem{Bt34} A. Niederberger, B. A. Malomed, and M. Lewenstein, 
                     {\rm Generation of optical and matter-wave solitons in binary 
                      systems with a periodically modulated coupling},
                      Phys. Rev. A {\bf 82}, 043622 (2010).

\bibitem{Bt36} S. S. Natu and E. J. Mueller,
                     {\rm Domain-wall dynamics in a two-component Bose-Mott insulator},
                     Phys. Rev. A {\bf 82}, 013612 (2010).

\bibitem{Bt37} P. Zi\'n, B. Ole\'s, and K. Sacha,
                     {\rm Quantum particle-number fluctuations 
                     in a two-component Bose gas in a double-well potential},
                     Phys. Rev. A {\bf 84}, 033614 (2011).

\bibitem{Bt38} K. Sasaki, N. Suzuki, and H. Saito, 
                     {\rm Dynamics of bubbles in a two-component Bose-Einstein condensate},
                     Phys. Rev. A {\bf 83}, 033602 (2011).

\bibitem{Bt39} B. Chatterjee, I. Brouzos, L. Cao, and P. Schmelcher,
                     {\rm Few-boson tunneling dynamics of strongly correlated binary mixtures in a double well},
                     Phys. Rev. A {\bf 85}, 013611 (2012).

\bibitem{Bt40} L. Cao, I. Brouzos, B. Chatterjee, and P. Schmelcher,
                     {\rm The impact of spatial correlation on the tunneling dynamics of few-boson 
                     mixtures in a combined triple well and harmonic trap},
                     New J. Phys. {\bf 14}, 093011 (2012).

\bibitem{Bt41} O. E. Alon, A. I. Streltsov, K. Sakmann, A. U. J. Lode, J. Grond, and L. S. Cederbaum,
                     {\rm Recursive formulation of the multiconfigurational time-dependent Hartree method 
                     for fermions, bosons and mixtures thereof in terms of one-body density operators},
                     Chem. Phys. {\bf 401}, 2 (2012).

\bibitem{Bt42} S. Kr\"onke, L. Cao, O. Vendrell, and P. Schmelcher,
                     {\rm Non-equilibrium quantum dynamics of ultra-cold atomic mixtures: 
                     the multi-layer multi-configuration time-dependent Hartree method for bosons},
                     New J. Phys. {\bf 15}, 063018 (2013).

\bibitem{Bt42h} K. Anoshkin, Z. Wu, and E. Zaremba,
                       {\rm Persistent currents in a bosonic mixture in the ring geometry},
                       Phys. Rev. A {\bf 88}, 013609 (2013).                       

\bibitem{Bt43} L. Cao, S. Kr\"onke, O. Vendrell, and P. Schmelcher,
                     {\rm The multi-layer multi-configuration time-dependent Hartree method for bosons: 
                     Theory, implementation, and applications},
                     J. Chem. Phys. {\bf 139}, 134103 (2013).

\bibitem{Bt44} B. Gertjerenken, T. P. Billam, C. L. Blackley, C. R. Le Sueur, L. Khaykovich, S. L. Cornish, and C. Weiss,
                     {\rm Generating Mesoscopic Bell States via Collisions of Distinguishable Quantum Bright Solitons},
                     Phys. Rev. Lett. {\bf 111}, 100406 (2013).

\bibitem{Bt45} M. A. Garc\'ia-March, B. Juli\'a-D\'iaz, G. E. Astrakharchik, Th. Busch, J. Boronat, and A. Polls, 
                     {\rm Quantum correlations and spatial localization in one-dimensional ultracold bosonic mixtures},
                     New J. Phys. {\bf 16} 103004 (2014).

\bibitem{Bt46} L. Cao, S. Kr\"onke, J. Stockhofe, J. Simonet, K. Sengstock, D.-S. L\"uhmann, and P. Schmelcher,
                     {\rm Beyond-mean-field study of a binary bosonic mixture in a state-dependent honeycomb lattice},
                     Phys. Rev. A {\bf 91}, 043639 (2015).

\bibitem{Bt47} S. Kr\"onke, J. Kn\"orzer, and P. Schmelcher,
                     {\rm Correlated quantum dynamics of a single atom collisionally coupled 
                     to an ultracold finite bosonic ensemble},
                     New J. Phys. {\bf 17}, 053001 (2015). 

\bibitem{Bt48} D.  S. Petrov,
                     {\rm Quantum Mechanical Stabilization of a Collapsing Bose-Bose Mixture},
                     Phys. Rev. Lett. {\bf 115}, 155302 (2015).
 
\bibitem{HIM_Cohen} L. Cohen and C. Lee,
                    {\rm Exact reduced density matrices for a model problem}, 
                    J. Math. Phys. {\bf 26}, 3105 (1985).

\bibitem{HIM_Po0} M. A. Za\l{}uska-Kotur, M. Gajda, A. Or\l{}owski, and J. Mostowski,
                           {\rm Soluble model of many interacting quantum particles in a trap},
                           Phys. Rev. A {\bf 61}, 033613 (2000).

\bibitem{HIM_Yan} J. Yan, 
                           {\rm Harmonic Interaction Model and Its Applications in Bose-Einstein Condensation},
                           J. Stat. Phys. {\bf 113}, 623 (2003).

\bibitem{HIM_Po1} M. Gajda, 
                           {\rm Criterion for Bose-Einstein condensation in a harmonic 
                           trap in the case with attractive interactions}, 
                           Phys. Rev. A {\bf 73}, 023603 (2006).

\bibitem{HIM_Benchmarks} A. U. J. Lode, K. Sakmann, O. E. Alon, L. S. Cederbaum, and A. I. Streltsov, 
                                      {\rm Numerically exact quantum dynamics of bosons with 
                                      time-dependent interactions of harmonic type},
                                      Phys. Rev. A {\bf 86}, 063606 (2012).

\bibitem{LR_MCTDHB_Bench} O. E. Alon, 
                                         {\rm Many-body excitation spectra of trapped bosons with general interaction by linear response},
                                         J. Phys.: Conf. Ser. {\bf 594}, 012039 (2015). 
 
\bibitem{Schilling_FER_HIM_WORK} C. Schilling and R. Schilling,
                                                  {\rm Number-parity effect for confined fermions in one dimension},
                                                  Phys. Rev. {\bf 93}, 021601(R) (2016).

\bibitem{Axel_MCTDHF_HIM} E. Fasshauer and A. U. J. Lode,
                                         {\rm Multiconfigurational time-dependent Hartree method for fermions: 
                                         Implementation, exactness, and few-fermion tunneling to open space},
                                         Phys. Rev. A {\bf 93}, 033635 (2016).

\bibitem{Lenz_Domcke_Harmonic} L. S. Cederbaum and W. Domcke,
                                               {\rm A many-body approach to the vibrational structure in molecular 
                                               electronic spectra. I. Theory},
                                               J. Chem. Phys. {\bf 64}, 603 (1976).

\bibitem{Yngvason_PRA} E. H. Lieb, R. Seiringer, and J. Yngvason,
                                  {\rm Bosons in a trap: A rigorous derivation of the Gross-Pitaevskii energy functional},
                                   Phys. Rev. A {\bf 61}, 043602 (2000).

\bibitem{Lieb_PRL} E. H. Lieb and R. Seiringer,
                           {\rm Proof of Bose-Einstein Condensation for Dilute Trapped Gases},
                           Phys. Rev. Lett. {\bf 88}, 170409 (2002).

\bibitem{Erdos_PRL} L. Erd\H{o}s, B. Schlein, and H.-T. Yau,
                             {\rm Rigorous Derivation of the Gross-Pitaevskii Equation}, 
                             Phys. Rev. Lett. {\bf 98}, 040404 (2007).

\bibitem{MATH_ERDOS_REF} L. Erd\H{o}s, B. Schlein, and H.-T. Yau,
                                            {\rm Derivation of the cubic non-linear Schr\"odinger equation 
                                            from quantum dynamics of many-body systems},
                                            Invent. Math. {\bf 167}, 515 (2007). 

\bibitem{Variance} S. Klaiman and O. E. Alon,
                          {\rm Variance as a sensitive probe of correlations}, 
                          Phys. Rev. A {\bf 91}, 063613 (2015).

\bibitem{TD_Variance} S. Klaiman, A. I. Streltsov, and O. E. Alon, 
                                {\rm Uncertainty product of an out-of-equilibrium many-particle system},
                                Phys. Rev. A {\bf 93}, 023605 (2016).

\bibitem{QM_book} C. Cohen-Tannoudji, B. Diu, and F. Lalo\"e, 
                           {\it Quantum Mechanics}, Vol. 1 
                           (Wiley, New York, 1977).

\bibitem{Oriol_Robin} Y.-J. Chen, S. Pabst, Z. Li, O. Vendrell, and R. Santra,
                               {\rm Dynamics of fluctuations in a quantum system},
                               Phys. Rev. A {\bf 89}, 052113 (2014). 
 
\bibitem{REMARK} 
            Note that the initial conditions entering the mixture's recursion, 
            $\tilde C_M=-\gamma$ and $D_M=\gamma$ in (\ref{FUN_N_HALF_M1_2}), 
            do not vanish whereas the initial condition entering the recurrence relation 
            of the HIM system does \cite{HIM_Cohen}.
            This, of course, makes the particular 
            solutions of the respective recursions different 
            (unless the inter-species interaction $\lambda_2$ vanishes, $\gamma=0$,
            and we have two independent Bose-Einstein condensates).
 
\end{thebibliography}
\end{document}